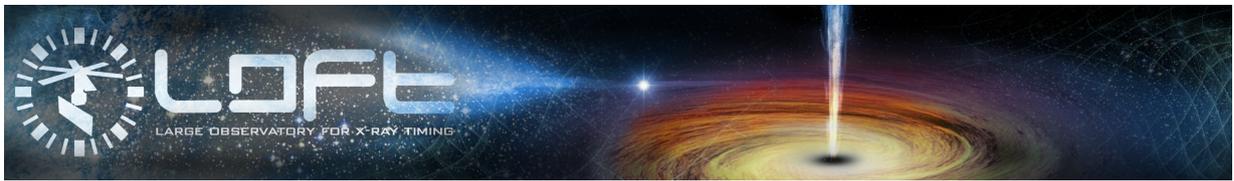

# *LOFT* as a discovery machine for jetted Tidal Disruption Events

White Paper in Support of the Mission Concept of the Large Observatory for X-ray Timing

Authors

E. M. Rossi[1], I. Donnarumma[2], R. Fender[3], P. Jonker[4,9],
S. Komossa[5], Z. Paragi[6], I. Prandoni[7], L. Zampieri[8]

[1] Leiden Observatory, Leiden University, P.O. Box 9513, 2300 RA , Leiden, The Netherlands
[2] INAF-IAPS, Via Fosso del Cavaliere 100, 00133, Rome, Italy
[3] Department of Physics, Oxford University, Keble Road, OX1 3RH, Oxford, UK
[4] Netherlands Institute of Space Research, , Netherlands Institute of Space Research, Utrecht, the Netherlands
[5] Max-Planck-Institut für Radioastronomie, Auf dem Hügel 69, 53121 Bonn, Germany
[6] Joint Institute for VLBI in Europe, Postbus 2, NL-7990 AA Dwingeloo, the Netherlands
[7] INAF-IRA Bologna, Via P. Gobetti 101, 40129 Bologna, Italy
[8] INAF-Osservaorio Astronomico di Padova, Vicolo dell'Osservatorio 5, I - 35122 Padova , Italy
[9] Department of Astrophysics/IMAPP, Radboud University Nijmegen, P.O. Box 9010, 6500 GL Nijmegen, The Netherlands

## Preamble

The Large Observatory for X-ray Timing, *LOFT*, is designed to perform fast X-ray timing and spectroscopy with uniquely large throughput (Feroci et al., 2014). *LOFT* focuses on two fundamental questions of ESA's Cosmic Vision Theme "Matter under extreme conditions": what is the equation of state of ultra-dense matter in neutron stars? Does matter orbiting close to the event horizon follow the predictions of general relativity? These goals are elaborated in the mission Yellow Book (http://sci.esa.int/loft/53447-loft-yellow-book/) describing the *LOFT* mission as proposed in M3, which closely resembles the *LOFT* mission now being proposed for M4.

The extensive assessment study of *LOFT* as ESA's M3 mission candidate demonstrates the high level of maturity and the technical feasibility of the mission, as well as the scientific importance of its unique core science goals. For this reason, the *LOFT* development has been continued, aiming at the new M4 launch opportunity, for which the M3 science goals have been confirmed. The unprecedentedly large effective area, large grasp, and spectroscopic capabilities of *LOFT*'s instruments make the mission capable of state-of-the-art science not only for its core science case, but also for many other open questions in astrophysics.

*LOFT*'s primary instrument is the Large Area Detector (LAD), a $8.5\,\text{m}^2$ instrument operating in the 2–30 keV energy range, which will revolutionise studies of Galactic and extragalactic X-ray sources down to their fundamental time scales. The mission also features a Wide Field Monitor (WFM), which in the 2–50 keV range simultaneously observes more than a third of the sky at any time, detecting objects down to mCrab fluxes and providing data with excellent timing and spectral resolution. Additionally, the mission is equipped with an on-board alert system for the detection and rapid broadcasting to the ground of celestial bright and fast outbursts of X-rays (particularly, Gamma-ray Bursts).

This paper is one of twelve White Papers that illustrate the unique potential of *LOFT* as an X-ray observatory in a variety of astrophysical fields in addition to the core science.

# 1 Introduction

Mounting observational evidence is supporting a scenario where most galactic nuclei host supermassive black holes (SMBHs). Gas inflow from larger scales causes a small (∼ 1%) fraction of SMBHs to accrete continuously for millions of years and shine as quasars. Most of them are instead "quiescent", accreting –if at all– at a highly sub-Eddington rate. Observationally, it is therefore hard to assess the presence and measure the mass of most of SMBHs, beyond the local universe. Occasionally, however, a sudden increase of the accretion rate may occur, if a large mass of gas, e.g. a star, falls into the tidal sphere of influence of the black hole and finds itself torn apart and accreted. We call these events "tidal disruption events" (TDEs). TDEs can result in a sudden rise of electromagnetic emission. They can reach the luminosity of a quasar but they are rare (∼ $10^{-5}$ yr$^{-1}$ per galaxy) and last several months, or years in *soft* X-ray Komossa (2002).

The detection and study of these flares can deliver other important astrophysical information, beyond probing the presence of a SMBH. After stellar disruption, part of the stellar material is accreted onto the black hole, causing a luminous flare of radiation. If the star is completely disrupted, its debris is accreted at a decreasing rate of $\dot{M} \propto t^{-5/3}$ (Rees, 1988; Phinney, 1989). Therefore, TDEs allow us to study the formation of a transient accretion disc and its continuous transition through different accretion states. The super-Eddington phase — which occurs only for SMBH masses $M \lesssim 10^7$ M$_\odot$ — is theoretically uncertain, but it may be associated with a powerful radiatively driven wind (Rossi & Begelman, 2009), that thermally emits ∼ $10^{41} - 10^{43}$ erg s$^{-1}$, mainly at optical frequencies (Strubbe & Quataert, 2009; Lodato & Rossi, 2011). The disc luminosity (∼ $10^{44} - 10^{46}$ erg s$^{-1}$), instead, peaks in the far-UV/soft X-rays (Lodato & Rossi, 2011). Of great theoretical importance would also be the possibility to observe the formation and evolution of an associated jet, powered by this sudden accretion episode. Furthermore, TDEs are signposts of supermassive binary black holes, as their lightcurves look characteristically different in the presence of a second BH, which acts as a perturber on the stellar stream (Liu et al., 2009), as recently observed in a first candidate event (Liu et al., 2014). Finally, we note that, in X-rays, TDEs represent a new probe of strong gravity, for instance tracing precession effects in the Kerr metric.

The first TDEs were discovered in *ROSAT* surveys of the X-ray sky (e.g., Komossa & Bade, 1999; Grupe et al., 1999) — see Komossa (2002) for a review. Later, *GALEX* allowed for the selection of TDEs at UV frequencies (Gezari et al., 2009; Gezari et al., 2012); many of the most recent TDE candidates are now found in optical transient surveys (Van Velzen et al., 2011a; Cenko et al., 2012; Chornock et al., 2014). An alternative method to select TDE candidates is to look for optical spectra with extreme high-ionization and Balmer lines (Komossa et al., 2008; Wang et al., 2012).This "thermal" Spectral Energy Distribution (SED) of TDEs is believed to be associated with emission from the disc or the radiatively driven wind.

Recently the *Swift* Burst Alert Telescope (BAT) instrument (Bloom et al., 2011; Burrows et al., 2011; Cenko et al., 2012), triggered two TDE candidates in the hard X-ray band. A multi-frequency follow-up from radio to γ-rays revealed a new class of *non-thermal* TDEs. It is widely believed that emission from a relativistic jet (Γ ≈ 2, Zauderer et al., 2013; Berger et al., 2012) is responsible for the hard X-ray spectrum (with power-law photon index α ∼ 1.6 − 1.8) and the increasing radio activity (Levan et al., 2011), detected a few days after the trigger. The best studied of the two events is Swift J1644+57 (Sw J1644 in short). The two main features that support the claim that Sw J1644 is a TDE are i) the X-ray lightcurve behaviour, that follows $L_x \propto \dot{M} \propto (t-\tau)^{-5/3}$ after a few days (τ ≈ 3 day) from the trigger, and ii) the radio localization of the event within 150 pc from the centre of a known quiescent galactic nucleus (Zauderer et al., 2011).

In this White Paper, we use the Sw J1644 X-ray lightcurve (see next section) to estimate detection rate prospects for the proposed M4 mission *LOFT* (Feroci et al., 2012). We chose to focus on predictions for non-thermal TDEs, best suited to be a *LOFT* target (*LOFT* could also see the late corona hard X-ray emission in disk-dominated TDEs but only in the local Universe; Strubbe & Quataert, 2009). On board there are two main instruments: the Large Area Detector (LAD, 2-50 keV) optimally suited for X-ray timing studied thanks to the 8.5 m$^2$ area peaking at ∼ 8 keV and the Wide Field Monitor (WFM) with a ∼ 4 sr FoV aimed at triggering follow-up observations with the LAD for a variety of sources (both galactic and extra-galactic). We consider both and give possible discovery and identification strategies. Our conclusions can be summarized as follows:

- the Wide Field Monitor can allow to serendipitously detect non-thermal TDEs, at a higher rates (between one and a few tens per year) than any currently planned survey missions.

- the Large Area Detector can effectively follow up and identify radio triggered candidates. In synergy with SKA in survey mode, LAD can in principle successfully repoint each radio candidate, and measure its lightcurve decay index and spectral properties. LAD, together with Athena, will be the best X-ray follow up instrument in the SKA era. The expected rate is between a few to a few hundreds per yr.

- *LOFT* is a unique mission for *jetted* TDEs, because it can combine the benefits of a WFM and a follow-up instrument with a wide energy range (up to 50 keV). The attractive prospect is that of building for the first time a sample of well studied TDEs associated with non-thermal emission from jets. This will allow us to study disc-jet formation and their connection, in a complementary way with respect to other more persistent sources such as active galactic nuclei and X-ray binaries. In addition, jetted TDEs can uniquely probe the presence of supermassive black hole in *quiescent* galaxies well beyond the local universe.

Throughout this paper we use the following cosmological parameters: $\Omega_M = 0.25$, $\Omega_\lambda = 0.75$ and $H_0 = 70$ km/s/Mpc.

## 2 $X$−ray emission of jetted TDEs

In 2011, BAT discovered what was first thought to be a peculiar gamma-ray burst, now known as Sw J1644. What first caught the attention was its flaring activity, which caused multiple BAT triggers, followed by an X-ray light curve that extended to much longer timescales than any known GRB (Levan et al., 2011). The source was confirmed to be of extra-galactic origin, with a peak luminosity ∼ 100 times higher than that of bright active galactic nuclei (Burrows et al., 2011; Levan et al., 2011). The observed 0.3-10 keV lightcurve of Sw J1644 is shown in Fig.1 (blue squares). Activity was already detected ≈ 3 days before the BAT "official" trigger and the start of XRT observations (Burrows et al., 2011). Considering this delay, the X-ray lightcurve as a function of time $\Delta t$ since the trigger ($\Delta t = 0$) may be described by

$$L_{x,iso}(\Delta t) \approx 1.5 \times 10^{48} \text{ erg s}^{-1} \left(\frac{\tau + \Delta t}{\tau}\right)^{-5/3}, \tag{1}$$

(Fig.1, solid line). $L_{x,iso}$ is the isotropic equivalent luminosity, computed from the X-ray flux. The X-ray luminosity can be related to the jet luminosity assuming a constant radiation efficiency in X-rays $\epsilon_x$, $L_j = L_{x,iso}(1 - \cos\theta_j)/(\epsilon_x \delta^2)$, where $\theta_j$ is the jet opening angle and $\delta$ the Doppler factor. The radiation efficiency of J1644 in 1-10 keV band (i.e. the fraction of the total luminosity emitted in that band) is inferred from BAT, XRT and Fermi spectral information to be $\epsilon_x \approx 0.2$ (Burrows et al., 2011). We note that the X-ray spectrum of Sw 1644 remained hard ($\alpha \sim 1.4 - 1.7$) over the duration of the event with the exception of the short dips in the initial variable phase (Saxton et al., 2012).

In the following, we assume that the lightcurve of J1644 is the TDE prototypical X-ray lightcurve in the 1-10 keV band, for events where we observe non-thermal and relativistically boosted emission, associated with jets. Then, if the jet luminosity traces the mass accretion rate, $L_j \propto \dot{M}$, it is possible to rescale eq.1 for events with different delay of trigger time, redshift, supermassive black hole (SMBH) and stellar mass (see Donnarumma & Rossi, 2014, for details).

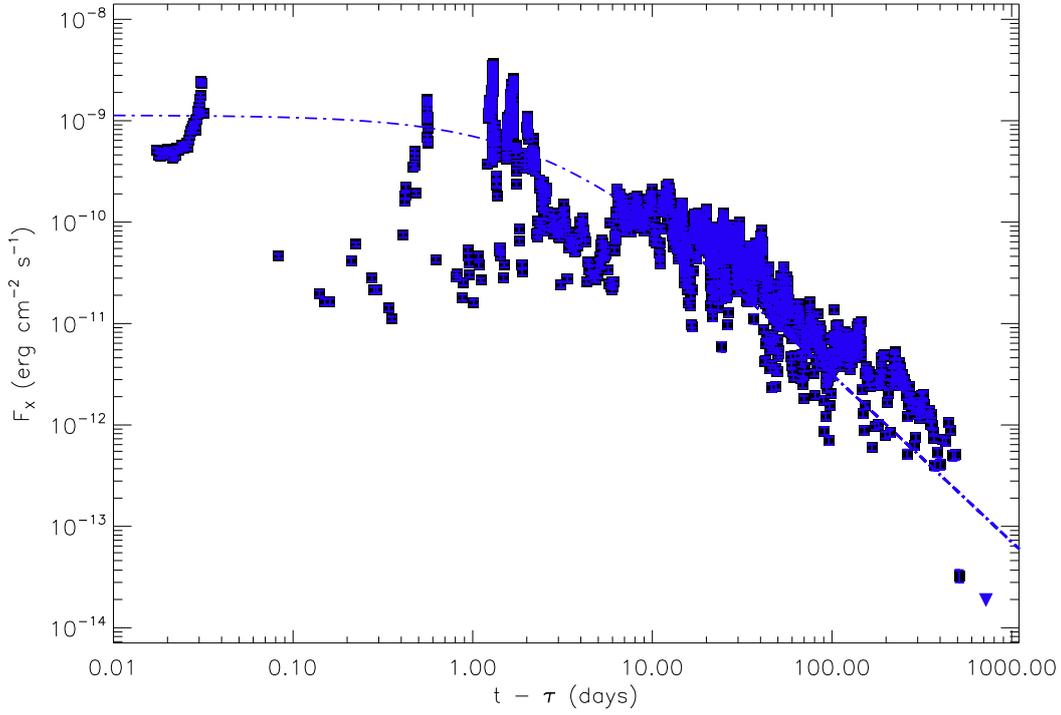

Figure 1: The X-ray (0.3-10 keV) lightcurve of J1644 as a function of time from the X-ray trigger: data (absorbed flux, blue squares taken from publicly available XRT lightcurves at http://www.swift.ac.uk/xrt_curves/) versus our modeling (dot-dashed line). The decay can be fitted by $(t-\tau)^{-5/3}$, with $\tau = 3$ day.

## 3 Monte Carlo calculations for rate predictions

We construct a population of TDEs, with each event characterized by a trigger time $\tau$, a redshift, a SMBH mass $M$ and a mass of the star $m_*$ that is disrupted. The trigger time is chosen randomly from a one year interval. The mass distribution of SMBHs as a function of redshift is taken from Shankar et al. (2013). In particular, we consider the two accretion models that yield the largest and the lowest mass function predictions, to quantify the uncertainty on our expected TDE rates, due to the imperfect knowledge of the black hole mass distribution. Finally, we assume a population of main sequence stars in galactic nuclei that follows a Kroupa Initial Mass Function, (Kroupa, 2001).

This intrinsic distribution is then rescaled by $2\pi\Gamma^{-2}/(4\pi) = 1/(2\Gamma^2) \approx 0.125$, i.e. the fraction of solid angle subtended by the emission, from a two sided jet. We are now in the position to make predictions for specific instruments, once an observing strategy is specified. In the following two sections, we focus on the instruments on board of *LOFT*. We then compare their performances with those of other planned missions in sec. 6.

## 4 Prospect for *direct* detection with *LOFT* Wide Field Monitor

The Wide Field Monitor (WFM) of *LOFT* (Feroci et al., 2012) will survey $1/3^{rd}$ of the sky in each pointing, reaching a 5-$\sigma$ sensitivity of $\sim 8-9 \times 10^{-11}$ erg cm$^{-2}$ s$^{-1}$ per day in the energy range 2-50 keV. The combination of its sky coverage and sensitivity makes the WFM a good hunter for *serendipitously* detecting TDEs. To be able to detect and identify a TDE, we adopt a daily flux threshold that corresponds roughly to a 25-$\sigma$ detection, allowing us to measure its temporal decaying index. We therefore start from the 1 day 5-$\sigma$ flux limit in $2-50$ keV band and then we extrapolate it backward using the expected $t^{-5/3}$ decay. We then take the flux corresponding to 4 points over this period and compute their average. This average flux defines our *identification* flux threshold. To claim identification, each event in our simulations needs to have a flux averaged over one day above this flux limit.

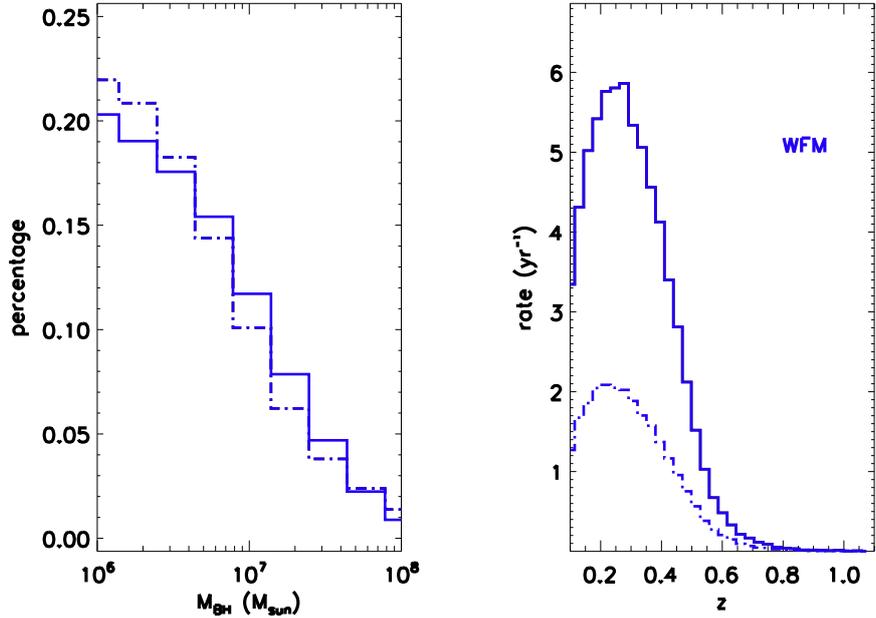

Figure 2: Left panel: percentage of event rates as a function of BH mass. Right panel: event rates for the WFM on board of *LOFT* as a function of redshift. In both panels, two different black hole distribution functions are shown (blue solid line: G model; blue dot-dashed line: Gz from Shankar et al. (2013)).

The results are shown in Fig.2 and Table 1. As showed in Fig. 2 (left panel), the TDE rate distribution steadily increases going to lower BH masses from a few percent at $M_{BH} \sim 10^8 M_\odot$ up to a maximum of $\sim 20\%$ at $M_{BH} \sim 10^6 M_\odot$. The rate distribution drop to a few percents at $M_{BH} \sim 10^8 M_\odot$ is a natural consequence of the BH mass dependence in Eq. 1. In total, we estimate tens of events per year up to $z_{max} \approx 0.6$, where we define $z_{max}$ as the redshift where the rate distribution reaches a value of 0.5 yr$^{-1}$. The peak rate is $\sim 6$ yr$^{-1}$ at $z = 0.2 - 0.3$ (see Figure 2, right panel). We note that the reported values are for a beaming factor of $\Gamma = 2$, as measured from radio observations. When comparing our results with BAT/Swift observations, however, one realizes that a suppression factor for these rates may be needed (see Donnarumma & Rossi, 2014, for details). The quantification of this factor is challenging because BAT does not operate in survey mode and there are over 500 trigger criteria, that make our use of a flux limit survey a rather simplified approach. A somewhat hard lower limit estimate suggests a suppression factor of two orders of magnitude. This can be due to a combination of a larger beaming factor and a jet production efficiency < 1. For example, a beaming factor of $\Gamma = 20$ in the X-ray emission region would account for the BAT rate and be consistent with high energy observation (Burrows et al., 2011). This would imply physically different radio and X-ray emitting regions. Therefore in the most pessimistic case, the WFM would observe TDEs at a rate of $\approx 0.3 - 0.8$ yr$^{-1}$, similar to that of BAT. For a lower limit on the rates for all X-ray instruments mentioned in the following, the same factor should be applied. Of course, this does note change the discussion on the relative performance of the missions.

## 5 LAD as excellent follow-up instrument of SKA triggers

The LAD on board of *LOFT* (2-50 keV) is a collimated instrument with 1 degree field of view, and a wide energy range (2-50 keV). Its background limited $10 - \sigma$ sensitivity is $\geq 3 \times 10^{-12}$ erg cm$^{-2}$ s$^{-1}$ in the 2-10 keV band, for a 1000 s exposure (see Fig.3).

Given the smaller field of view and the higher sensitivity with respect to the WFM, LAD is more effective as a follow-up instrument. Jetted TDEs have a strong radio counterpart and the Square Kilometer Array (SKA Dewdney et al., 2013) in survey mode has the potential to detect a few hundred events per year Van Velzen et al. (2011); Donnarumma & Rossi (2014). We therefore investigate a possible strategy to efficiently following-up SKA triggered candidate events and then assess the actual rate of their X-ray counterparts. Since the radio TDE emission is quite featureless and the TDE signature (i.e. the $t^{-5/3}$ decay) seems to be a prerogative of its X-ray emission, the X-ray follow-up of radio candidates is essential for their *identification*, together with a precise radio localisation. The role of *LOFT* in this respect becomes fundamental because Swift, Chandra and XMM are unlikely to be operating by the time SKA in survey mode will be detecting candidates.

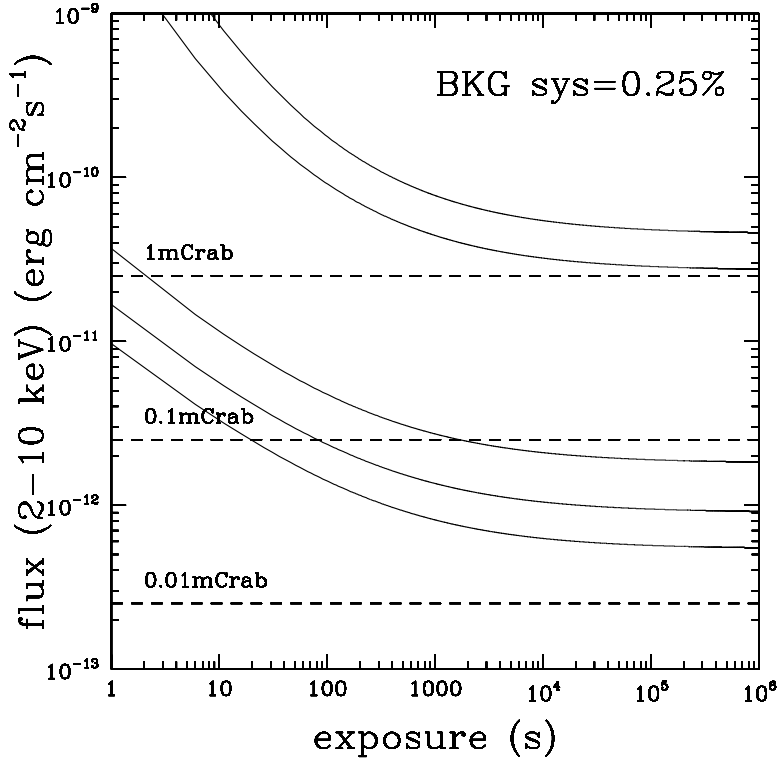

Figure 3: LAD sensitivity limit as a function of exposure time for different S/N ratios corresponding to a systematics of 0.25% on the background knowledge. From bottom to top: S/N~ 3, 5, 10, 150, 250.

We consider a 1-day delay in the X-ray repointing. Similarly to the strategy adopted for the WFM, we require at least 4 pointings spread over 4 days, each with a S/N ratio of at least 10. Each of them requires an exposure time less than 1000 s. This high S/N is required in order to characterize both the temporal and spectral behavior of the source. Following the same procedure explained in the previous section, we define an identification flux threshold of $\geq 10^{-11}$ erg cm$^{-2}$ s$^{-1}$ in the 2-10 keV band, that is then translated in the corresponding un-absorbed value in the 1-10 keV band, the energy range adopted in our modelling.

For each event in the Monte Carlo simulations, we calculate the average X-ray flux over the 4 pointings and compare it with the identification flux threshold. Finally, our detection rates are rescaled according to the fact that the *LOFT* repointing visibility allows a sky accessibility of ~ 75% (Feroci et al., 2014).

Figure 4 shows the expected rate of jetted TDEs as a function of BH mass (left panel) and of redshift (right panel). As regards their BH mass distribution, ~ 25% of all events have BHs with masses $\approx 10^6$ M$_\odot$, and events with BH masses $< 10^7$ M$_\odot$ dominate the redshift distribution at all epochs.

This time, the redshift distribution extends above $z = 1$ ($z_{max} \approx 1.2$) (see Table 1, bottom section), with most of the TDEs expected around $z \simeq 0.4$ (right panels). The peak rates are roughly between $\approx 4$ and 20 events per year. In total, LAD should be able to detect roughly half of the SKA triggered TDEs, i.e. between $\approx 130$ and $\simeq 350$ yr$^{-1}$. For these events, the instrument broad energy band $(2 - 50$ keV) should enable us to better characterize the energy budget of the X-ray component.

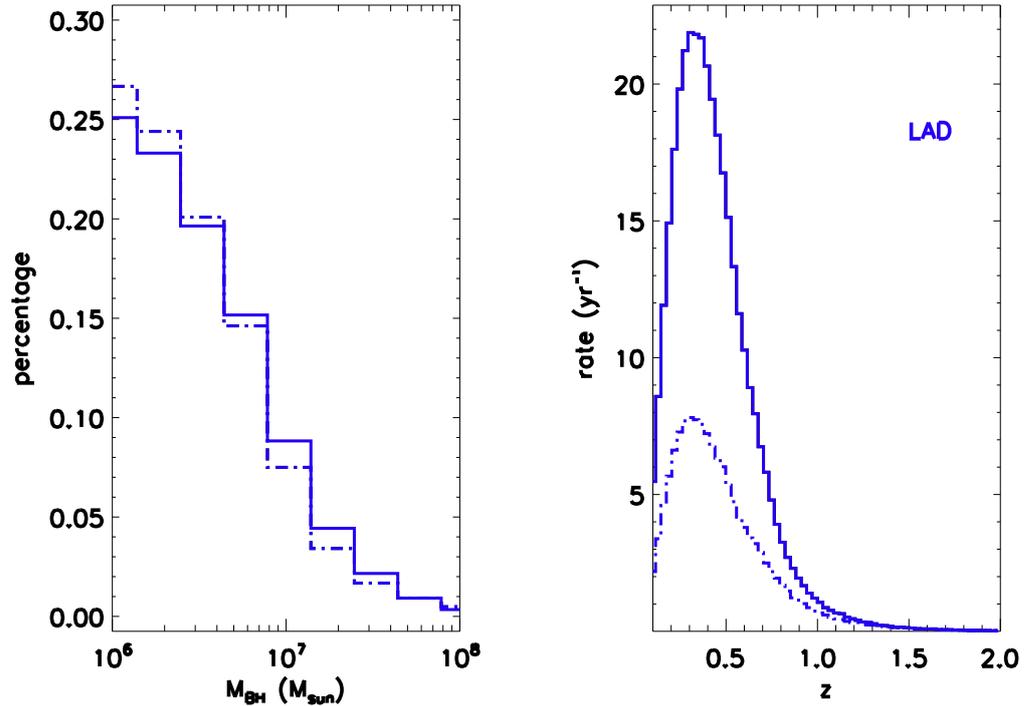

Figure 4: The same as Fig.2, but for the LAD instrument. These events have been first detected by SKA in wide radio surveys at 1.4 GHz. For details on the radio lightcurve modelling see Donnarumma & Rossi (2014), and in particular their model "MLD".

## 6 Discussion and conclusions

Although the LAD instrument would be roughly 5 times more effective in identifying TDEs, the WFM has the ability to serendipitously catch events, closer to their onset. This is shown in Fig.5, where the cumulative distribution of the trigger delay is shown as a function of redshift (from z=0.1 up to 0.6, that is the $z_{max}$ of WFM TDEs). While the WFM will be able to detect TDEs at any epoch, the LAD will preferentially catch them at later times, after $\approx 10$ days from their beginning. This latter delay is *not* due to an intrinsic problem of the LAD instrument, but it is a natural consequence of the delay introduced by the SKA trigger: the radio flux increases over time, and becomes bright enough for a SKA detection only after at least $\approx 10$ days from the beginning of the event (Donnarumma & Rossi, 2014). Another difference in the performance of the two instruments is the ability of LAD to go twice as far in redshift ($z \sim 1.2$) than WFM ($z \sim 0.6$). Synergy between the two instruments can also been envisaged. The LAD instrument can repoint within < 1 day candidates detected with WFM, allowing the study of the event lightcurve, also at early times (unlike SKA triggered TDEs).

The potential of the WFM for jetted TDE discoveries can be appreciated better comparing its performance with eRosita. Currently, this survey rapresents the best chance to serendipitously discover TDEs in X-ray, in the near future (Merloni et al., 2012). Indeed, the combination of a great sensitivity in 0.2 - 12 keV band and the pointing strategy (4-year all sky survey) makes it the optimal hunter of *thermal* TDEs, with ~ 1000 objects per scan at a sensitivity level several times better than RASS. However, it will be less performing at detecting non-thermal TDEs. Following the identification strategy detailed above, eRosita will be able to detect a factor 3 to 4 *less* jetted TDEs than the WFM in the hard X-ray band (see also Khabibullin et al., 2014). This is because eRosita will not cover an event repeatedly, but only every 6-month.

LAD can instead be compared with the Wide Field Imager (WFI) on board of Athena, the ESA L2 mission, planned to operate in 2028. We stress that radio counterparts need an X-ray follow-up to be confirmed and *LOFT*, together with Athena, is likely to be the only effective instrument operating simultaneously with SKA. If we use the very same follow-up strategy and we consider an optimistic repointing efficiency of 50%, then Athena may be able to identify slightly more TDEs (a factor 1.1-1.2). The reason of this small difference in identification rates is that only a moderate X-ray sensitivity ($\gtrsim 5 \times 10^{-12}$ erg cm$^{-2}$ s$^{-1}$) is required to obtain a complete set of X-ray counterparts of the radio triggered TDEs. However, the LAD wider energy range has the advantage of better constraining the total emitted luminosity and thus the jet energy content. Most importantly, Athena will not have a wide field monitor instrument on board and without a radio trigger will be blind to TDEs, while *LOFT* on the other hand can do TDE science autonomously. Thus, in this respect, *LOFT* is a more complete mission, combining the potential for serendipitous discoveries with the WFM and repointing ability with LAD.

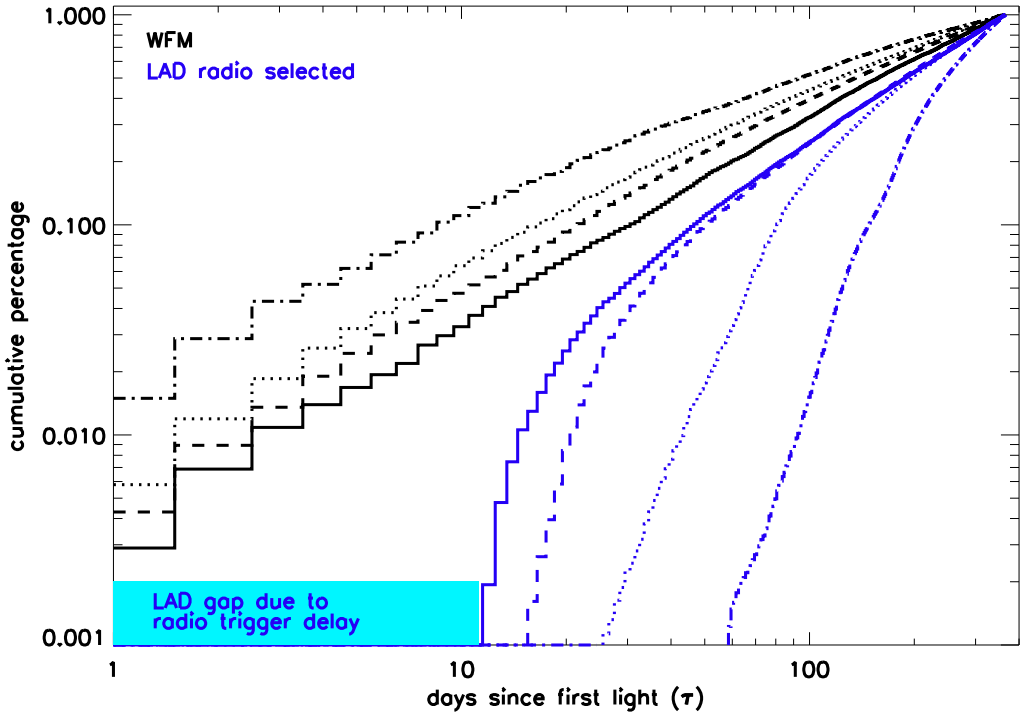

Figure 5: Cumulative distributions of delays in detecting the TDE from the explosion time, for different redshifts. We show results for the WFM (black lines) and LAD follow-ups of radio triggered TDEs (blue lines). The different line styles are for $z = 0.1, 0.2, 0.37, 0.6$ from right to left for WFM and vice versa for LAD. Note that the LAD instrument can repoint events only after 10 days, because the radio lightcurve becomes sufficiently bright to be detected by SKA only after that time.

In summary, *LOFT* has the potential to finally open a new era of statistical studies of TDEs and exploit them to address open questions in galaxy formation and high energy astrophysics. With a few tens to a few hundreds of events and their radio counterparts, one may constrain the SMBH mass function below redshift 1.5 and the physics of jet formation during transient accretion episodes: e.g. jet efficiency, Lorentz factor, and what is the connection with a super-Eddington accretion phase.

|  | $R^1$ yr$^{-1}$ | $R^2$ yr$^{-1}$ | $z_{\text{peak}}$ | $R^1_{\text{peak}}$ yr$^{-1}$ | $R^2_{\text{peak}}$ yr$^{-1}$ | $z_{\text{max}}$ |
|---|---|---|---|---|---|---|
| X-ray surveys |  |  |  |  |  |  |
| *LOFT* WFM | 24.5 | 67 | 0.2-0.3 | 2.3 | 6 | 0.6 |
| eRosita | 8 | 15 | 0.4 | 0.15 | 0.5 | 0.4 |
| Radio selected samples |  |  |  |  |  |  |
| *LOFT*LAD | 135 | 352 | 0.4 | 8 | 22 | 1.2 |
| Athena | 163 | 385 | 0.4 | 7 | 20 | 1.4 |

Table 1: Future X-ray surveys predictions: 1st and 2nd columns are total yearly rate (the subscripts $^1$ and $^2$ are for the *Gz* and *G* MFs, respectively), 3rd column redshift at the peak rate, 4th and 5th columns maximum peak rate and 6th column maximum redshift, defined as the z where the rate is 0.5. X-ray expected rates are derived for $\Gamma = 2$.